# Growth rate of the ancient human population in Australia

Ron W Nielsen aka Jan Nurzynski[1]

Environmental Futures Centre, Gold Coast Campus, Griffith University, Qld, 4222, Australia

September, 2013

Empirical growth rates of the ancient human population in Australia have been calculated using the data for the number of rock-shelter sites. They confirm the earlier conclusion that there was no intensification of growth between 10,000 and 1000 years BP.

**Introduction**

In the earlier publications (Nielsen aka Nurzynski, 2013a, 2013b) we have discussed the growth of human population in Australia as determined by the study of the time-dependent distribution of rock shelter sites between 10,000 and 1,000 years BP (before present). We have shown that if the number of rock shelter sites is assumed to represent also the size of human population, the growth of the population follows closely the second-order hyperbolic distribution. We have also shown that the analysis of rock-shelter data suggested a possible systematic error in relating the number of rock-shelter sites to the size of human population.

---

[1] r.nielsen@griffith.edu.au; ronwnielsen@gmail.com; http://home.iprimus.com.au/nielsens/ronnielsen.html

If the time-dependent distribution of the size of human population is corrected for this possible systematic error the growth of human population in Australia can be described using a simpler, first-order hyperbolic distribution. In any case, whether corrected or uncorrected, the general trend of growth of the ancient population in Australia is in good agreement with the trend of growth of the global human population (Johansen & Sornette, 2001; Karev, 2005; Korotayev, 2005; Shklovskii, 2002; von Foerster, Mora & Amiot 1960; von Hoerner, 1975).

We have pointed out that no claim is made that the "corrected" distribution is better. Both distributions, corrected and uncorrected, are suitable for studying general features of human population dynamics but not to claim the precise absolute values of the size of human population in Australia.

We have demonstrated that contrary to the illusion created by such distributions, the illusion reflected in the published claim about the intensification of growth around 5,000 years BP (Johnson & Brook, 2011), there was no intensification at any time between 10,000 and 1,000 years BP. Our aim now is to complement our previous discussion by the analysis of the empirical growth rates.

The growth rate $R(t)$ is defined as

$$R(t) = \frac{1}{S(t)} \frac{dS(t)}{dt} \qquad (1)$$

where $S(t)$ is the size of a growing entity.

For good-quality data, growth rate can be calculated directly from data. For poor quality data separated by large time intervals, empirical growth rate can be determined by using interpolated gradient divided by the relevant empirical size of the growing entity at a given time. This method was used to calculate empirical growth rates for the growth of human population in Australia between 10,000 and 1000 BP. Results are presented in Fig. 1.

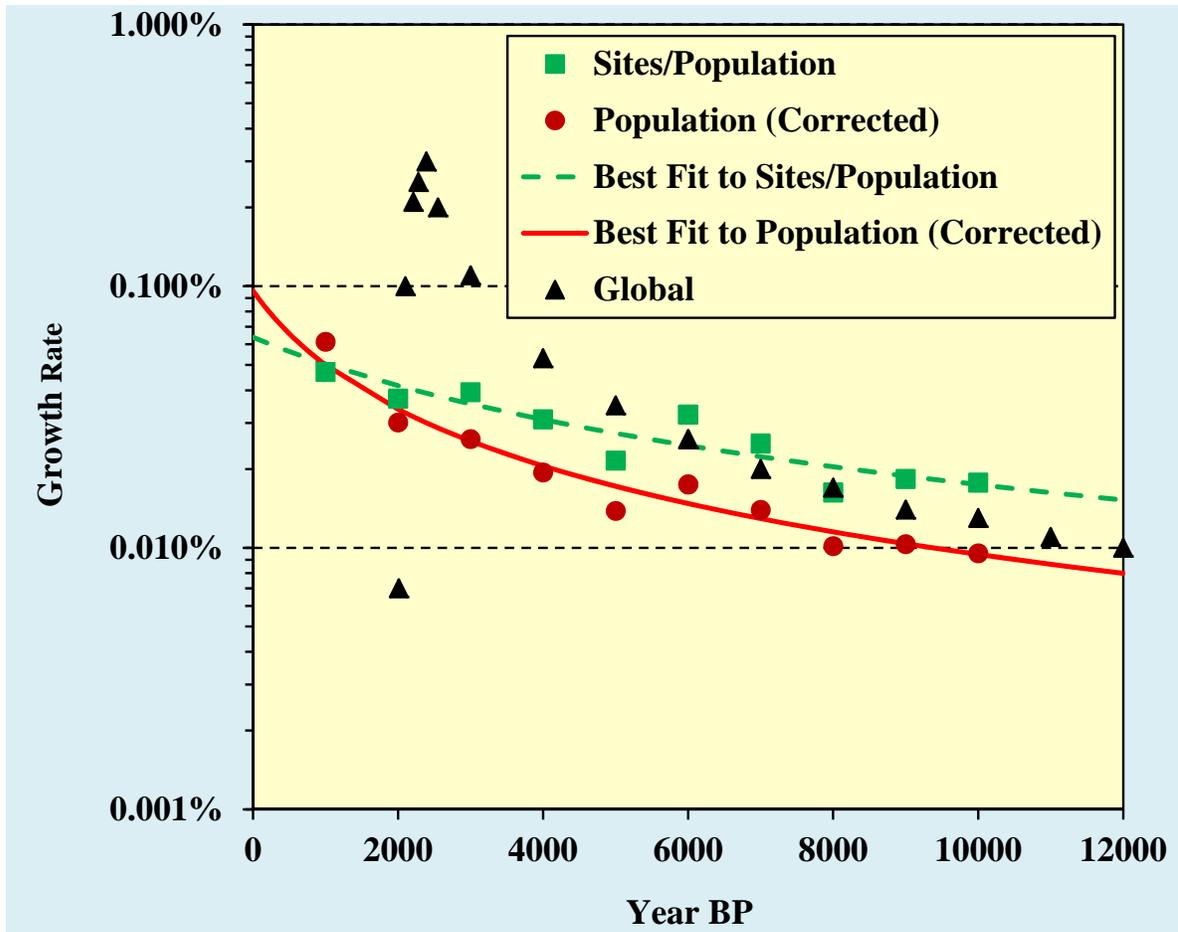

**Fig. 1.** Empirical growth rates of the ancient population in Australia between 10,000 and 1000 years BP are compared with the growth rates of the global population.

Empirical growth rates for the uncorrected and corrected sizes of the population follow monotonically increasing distributions (the arrow of time is from right to left), showing no sign of any intensification. Their values are comparable with the respective growth rates of the world population calculated using the best available data (Manning, 2008; US Census Bureau, 2013). The growth rate of the world population reached a maximum in 2390 years BP and commenced its rapid decline in agreement. Around 2000 years BP, the growth rate of the world population passed its zero value, which corresponds to a maximum in the growth of the world population at that time, and became negative.

The best fit to the empirical growth rate for Australia is obtained using the simple, first-order hyperbolic distribution:

$$N(t) = (a_0 + a_1 t)^{-1} \qquad (2)$$

where time $t$ is expressed in years BP, $a_0 = 1568.06$ and $a_1 = 0.4168$ are for the uncorrected growth of human population in Australia between 10,000 and 1,000 years BP while $a_0 = 1041.80$ and $a_1 = 0.9565$ are for the corrected growth.

The growth rate of the world population increases faster with time than the growth rate of the population in Australia. The gradient of the corrected distribution is in a slightly better agreement with the gradient for the world population but the absolute values do not allow for a positive discrimination between corrected and uncorrected trends. The absolute values for the growth rate of human population in Australia are in generally good agreement with the absolute values for the growth of the world population between around 10,000 and 5,000 years BP.

Empirical growth rates for the growth of ancient human population in Australia do not support the conclusion of Johnson and Brook (2011) that there was intensification of growth around 5000 years BP or at any other time between 10,000 and 1000 years BP. They do not support their claim that the growth of human population in Australia can be divided into two distinctly different stages "slow or negligible before 5000 years BP, and faster since then" (Johnson & Brook 2011, p. 1752). The growth rate was steadily increasing over the entire range of time between 10,000 and 1,000 years BP. Conclusion based on the investigation of empirical growth rates are in perfect agreement with our earlier conclusion (Nielsen aka Nurzynski, 2013a, 2013b) that nothing "important happened to the human population of Australia during the Holocene, and that the Mid-Holocene" and that there was no "turning point in Australian prehistory" (Johnson & Brook 2011, p. 1753), at least no turning point for the growth of human population within this range of time.